\documentclass[preprint,12pt]{article}
\usepackage{amssymb}
\usepackage{amsmath}
\usepackage{hyperref}

\begin{document}

\centerline{}

\centerline {\Large{\bf The nonlocal Darboux transformation  
 }}

\centerline{}

\centerline{\Large{\bf of the stationary axially symmetric Schr\"odinger equation  }}

\centerline{}

\centerline{\Large{\bf and generalized Moutard transformation}}

\centerline{}

\centerline{\bf {A. G. Kudryavtsev}}

\centerline{}

\centerline{Institute of Applied Mechanics,}

\centerline{Russian Academy of Sciences, Moscow 125040, Russia}

\centerline{kudryavtsev\_a\_g@mail.ru}

\begin{abstract}

The nonlocal Darboux transformation of the stationary axially symmetric 
Schr\"odinger equation is considered. It is shown that a special case of 
the nonlocal Darboux transformation provides the generalization of 
the Moutard transformation. Formulae for the generalized 
Moutard transformation are obtained. New examples of 
two - dimensional  potencials and exact solutions
for the stationary axially symmetric Schr\"odinger equation 
are obtained as an application of the generalized Moutard transformation.

\end{abstract}

\centerline{PACS: 02.30.Jr, 02.30.Ik, 03.65.Ge}






\section{Introduction}

Consider the stationary Schr\"odinger equation in the form
\begin{equation*}
\left( \Delta -{\it u}\left( x,y,z \right) \right) Y \left( x,y,z
\right) =0
\end{equation*}
In the case ${\it u}=-E+V\left( x,y,z \right)$ this equation 
describes nonrelativistic quantum system with energy $E$ \cite
{Landau}. In the case ${\it u}= -{\omega} ^{2}/{c\left( x,y,z
\right)}^{2}$ equation describes an acoustic pressure
field with temporal frequency $\omega$ in inhomogeneous media with
sound velocity $c$ \cite {Morse} and is known as Helmholtz equation
 \cite {Vinogradova}.  The case of fixed frequency
$\omega$ is of interest for modelling in acoustic tomography \cite
{Kak}. The case of fixed energy $E$ for two - dimensional equation 
is of interest for the multidimentional inverse scattering theory 
due to connections with two - dimensional integrable nonlinear 
systems \cite {Veselov1984}, \cite {Novikov2010}.

In the case of axial symmetry  the stationary Schr\"odinger equation 
 in cylindrical coordinates has the form

\begin{equation} \label{eq1}
 \left({\frac {\partial ^{2}}{\partial {r}^{2}}} +{\frac {1}{r}} 
{\frac {\partial }{\partial r}}+{\frac {
\partial ^{2}}{\partial {z}^{2}}}-u \left( r,z
 \right)  \right)Y \left( r,z \right) = 0
\end{equation}
The useful tool for one - dimensional Schr\"odinger equation is
the Darboux transformation \cite {Matveev1991}. The Moutard 
transformation  \cite {Moutard1878} is the two - dimensional 
generalization of the  Darboux transformation for the 
Schr\"odinger equation in cartesian coordinates. In the papers 
\cite {Kudryavtsev2013},  \cite {Kudryavtsev2016} the nonlocal 
Darboux transformation of the two - dimensional
stationary Schr\"odinger equation in cartesian coordinates was 
considered and its relation to the Moutard transformation was 
established. In the present paper we consider the stationary 
Schr\"odinger equation in cylindrical coordinates \eqref{eq1}
using the approach of papers \cite {Kudryavtsev2013},  
\cite {Kudryavtsev2016}.

We use the relation of the Schr\"odinger equation with the 
Fokker-Planck equation  \cite {Risken1989}. By the substitution
\begin{equation} \label{eq2}
Y \left( r,z \right) =W \left( r,z \right) {e^{h \left( r,z
\right) }}
\end{equation}
we obtain the equation
\begin{equation} \label{eq3}
{\frac {\partial }{\partial r}} \left( {\it W_{r}}
+2\,{ \it h_{r}}\,W+{\frac {1}{r}}\,W \right)+{\frac {\partial }{\partial z}} \left(
{\it W_{z}} +2\,{\it h_{z}}\,W \right) =0
\end{equation}
if $u$ and $h$ satisfy the condition
\begin{equation} \label{eq4}
{\it u}=-{\it h_{rr}} +{{\it h_{r}}}^{2}+{\frac {1}{r}}\, {\it h_{r}}+{\frac {1}{{r}^{2}}} 
-{\it h_{zz}} +{{\it h_{z}}}^{2}
\end{equation}

The equation \eqref{eq3} has the conservation low
form that yields a pair of equations

\begin{equation} \label{eq5}
{\it W_{r}}+ 2\,{ \it h_{r}}\,W+{\frac {1}{r}}\,W  -{\it Q_{z}} =0
\end{equation}
\begin{equation} \label{eq6}
{\it W_{z}}+ 2\,{ \it h_{z}}\,W  +{\it Q_{r}} =0
\end{equation}

As the variable $Q$ is a nonlocal variable for the Schr\"odinger 
equation \eqref{eq1} the Darboux transformation for the equations \eqref{eq5},
\eqref{eq6} provides the nonlocal Darboux transformation for the 
Schr\"odinger equation  \eqref{eq1}. 

\section{The nonlocal Darboux transformation 
and the generalization of the Moutard transformation}

Let us consider linear operator corresponding to the system of
equations \eqref{eq5}, \eqref{eq6}

\begin{equation*}
\hat L\left( h \left( r ,z \right)  \right) \, {\bf F}=
\begin{pmatrix} { 2\,{ \it h_{r}}+{\frac {1}{r}}+\frac {\partial }{\partial r} }  &
{ -\frac {\partial }{\partial z} } \\ {{2\,{ \it h_{z}}+\frac
{\partial }{\partial z}}}  & {\frac {\partial }{\partial r}}
\end{pmatrix}\,
\begin{pmatrix} F_1 \\ F_2 \end{pmatrix}
\end{equation*}

Consider Darboux transformation in the form

\begin{equation*}
\hat L_D \, {\bf F}=
\begin{pmatrix} { g_{11}-a_{11}\,\frac {\partial }{\partial r}-b_{11}\,\frac {\partial }{\partial z}  }  &
{  g_{12}-a_{12}\,\frac {\partial }{\partial r}-b_{12}\,\frac
{\partial }{\partial z} } \\ { g_{21}-a_{21}\,\frac {\partial
}{\partial r}-b_{21}\,\frac {\partial }{\partial z} }  & {
g_{22}-a_{22}\,\frac {\partial }{\partial r}-b_{22}\,\frac
{\partial }{\partial z} }
\end{pmatrix} \,
\begin{pmatrix} F_1 \\ F_2 \end{pmatrix}
\end{equation*}

If linear operators $\hat L$ and $\hat L_{D}$ hold the
intertwining relation

\begin{equation} \label{eq7}
\left( \hat L\left( h \left( r ,z \right) + s \left( r ,z
\right) \right)\hat L_{D} - \hat L_{D} \hat L\left( h \left( r ,z
\right) \right) \right) \, {\bf F}= 0
\end{equation}
for any $ {\bf F} \in  \mathcal{F} \supset Ker\left( \hat L\left(
h \right)\right)$ where  $Ker\left( \hat L\left( h
\right)\right)=\{{\bf F}:{\hat {L}}\left( h \right){\bf F}=0\}$,
then for any ${\bf F_s}\in Ker\left( \hat L\left( h
\right)\right)$
 the function $\tilde {\bf F} \left( r,z \right)=
\hat L_{D} {\bf F_s} \left( r,z \right)$ is a solution of the
equation ${\hat {L}}\left( {\tilde {h}} \right) \tilde {\bf F} =0$
\, with new potential $\tilde h = h+s$.

In accordance with the idea of the paper \cite {Kudryavtsev2013} 
we define domain ${\mathcal{F}}$ by restriction to solutions of 
one equation from the system of two equations \eqref{eq5}, \eqref{eq6}.
Consider equation \eqref{eq7} on the following domain:
\begin{equation*}
{\mathcal{F}}_0=\{\,{\bf F}: {F_1}_{r}+ 2\,{ \it h_{r}}\, F_1
+{\frac {1}{r}}\, F_1  - {F_2}_{z} =0\}
\end{equation*}

Taking into account this dependance of $F_1, F_2$ derivatives, the
equations for $s, g_{ij}, a_{ij}, b_{ij}$ can be obtained.

When solving equation \eqref{eq7} on the domain ${\mathcal{F}}_0$
the special situation arise in the case $s \left( r,z \right) = 
-2\,h \left( r,z \right) -\ln  \left( r \right)$. In this case we obtain 
from the equation \eqref{eq7} the following Darboux transformation

\begin{equation} \label{eq8}
\hat L_D = r\,{e^{2\,h \left( r,z \right) }}
\begin{pmatrix} { 0 }  &
{ 1 } \\ { - 1 }  & { 0 }
\end{pmatrix} \,
\end{equation}

By the formula \eqref{eq4} with $\tilde h = -h  -\ln  \left( r \right)$ we 
obtain for the new Schr\"odinger potential

\begin{equation} \label{eq9}
\tilde {u} \left( r ,z \right) = u\left( r ,z \right) 
+2\,{\frac {\partial ^{2}}{\partial {r}^{2}}}h \left( r,z \right) 
+2\,{\frac {\partial ^{2}}{\partial {z}^{2}}}h \left( r,z \right) 
-{\frac {1}{{r}^{2}}} 
\end{equation}

Consider
\begin{equation} \label{eq10}
Y_h\left( r ,z \right)={\frac {1}{r}}\,{{\rm e}^{-h \left( r,z \right) }}
\end{equation}

According to the formula \eqref{eq4}, $Y_h$ is a solution of the
Schr\"odinger equation  \eqref{eq1} with potential $u$.

Then we get another form of the formula \eqref{eq9}

\begin{equation} \label{eq11}
\tilde {u} \left( r ,z \right) = u\left( r ,z \right) 
-2\,{\frac {\partial ^{2}}{\partial {r}^{2}}}\ln  \left( {\it Y_h}
 \left( r,z \right)  \right) +{\frac {1}{{r}^{2}}} 
-2\,{\frac {\partial ^{2}}{\partial {z}^{2}}}\ln  
\left( {\it Y_h} \left( r,z \right)  \right) 
\end{equation}

This formula is the generalization of the formula 
of Moutard transformation for the potential of the 
Schr\"odinger equation.

From the formula \eqref{eq8} and relation

\begin{equation} \label{eq12}
\tilde {Y} \left( r,z \right) =\tilde {W} \left( r,z \right) {e^{
\tilde {h} \left( r,z \right) }}
\end{equation}

we have for the following equations

\begin{equation} \label{eq13}
\tilde {W} \left( r ,z \right) = r\,{e^{2\,h \left( r,z \right) }}Q
\left( r,z \right)
\end{equation}

\begin{equation} \label{eq14}
\tilde {Y} \left( r,z \right) = {e^{h \left( r,z \right) }}Q
\left( r,z \right)
\end{equation}

One can express $W, Q, h$ by equations \eqref{eq2}, \eqref{eq14},
\eqref{eq10} trough $Y, \tilde {Y}, Y_h$ and substitute to the
system of equations \eqref{eq5}, \eqref{eq6}. The result is

\begin{equation} \label{eq15}
{\frac {\partial }{\partial z}} \left( {\it Y_h} \left( r,z
\right) {\tilde {Y}} \left( r,z \right)  \right) - \left( {\it
Y_h} \left( r,z \right)  \right) ^{2}{\frac {\partial }{\partial
r}} \left( {\frac {{ \it Y} \left( r,z \right) }{{\it Y_h} \left(
r,z \right) }} \right)=0
\end{equation}

\begin{multline} \label{eq16}
{\frac {\partial }{\partial r}} \left( {\it Y_h} \left( r,z
\right) {\tilde {Y}} \left( r,z \right)  \right) 
+{\frac {1}{r}}\,{\it Y_h} \left( r,z \right) {\tilde {Y}} \left( r,z \right)
\\
+ \left( {\it Y_h} \left( r,z \right)  \right) ^{2}{\frac {\partial }{\partial
z}} \left( {\frac {{ \it Y} \left( r,z \right) }{{\it Y_h} \left(
r,z \right) }} \right)=0
\end{multline}

These formulae are the generalization of the formulae of the Moutard
transformation for the solution of the Schr\"odinger equation.

Thus the case $\tilde h = -h  -\ln  \left( r \right)$ of the 
nonlocal Darboux transformation for the stationary axially symmetric 
Schr\"odinger equation provides the generalization
of the Moutard transformation .

Note that 
${\it Y}={\it Y_h}, \, {\tilde {Y}}={\left(r\,{\it Y_h} \right)}^{-1} $
is the simple example of solution for equations 
\eqref{eq15}, \eqref{eq16}.

\section{Application of the generalized 
Moutard transformation }

For the first example of  of the generalized Moutard transformation 
application let us consider 
$u=0, {\it Y_h}={r}^{2}-2\,{z}^{2}, { \it Y}=z$.
From the equations \eqref{eq11}, \eqref{eq15}, \eqref{eq16} we
obtain potencial
\begin{equation} \label{eq17}
{\tilde {u}}_{1}  =
{\frac {4\,{z}^{4}+13\,{r}^{4}+20\,{r}^{2}{z}^{2}}{ \left( {r}^{2}-2\,
{z}^{2} \right) ^{2}{r}^{2}}}
\end{equation}
and solution of equation \eqref{eq1} with this potential
\begin{equation*}
{\tilde {Y}}_{1} =
{\frac {4\,{r}^{2}{z}^{2}+{r}^{4}+C_1}{r \left( {r}^{2}-2\,{z}^{2}
 \right) }}
\end{equation*}
where $C_1$ is an arbitrary constant. The potencial \eqref{eq17} has 
singularities where its denominator vanishes.

In the case of the two - dimensional Schr\"odinger equation 
in cartesian coordinates the twofold application of the Moutard 
transformation can be effective for obtaining nonsingular 
potentials \cite {Tsarev2008}, \cite {Kudryavtsev2016}.
Let us continue our example and make the second 
generalized Moutard transformation. Consider
$u={\tilde {u}}_{1}, {\it Y_h}={\tilde {Y}}_{1}$.
From the equation \eqref{eq11} we obtain potencial 
\begin{equation} \label{eq18}
{\tilde {\tilde {u}}}_{1}  =
{\frac {-8\,{r}^{2} \left(  \left( {r}^{2}-5\,{z}^{2} \right) ^{2}-33
\,{z}^{4} \right) -8\,C_1 \left( 5\,{r}^{2}+2\,{z}^{2} \right) }{
 \left( 4\,{r}^{2}{z}^{2}+{r}^{4}+C_1 \right) ^{2}}}\, .
\end{equation}
This potential is nonsingular if $C_1 > 0$.

To illustrate the obtaining of solutions for the 
stationary axially symmetric Schr\"odinger equation 
by the generalized Moutard transformation let us
derive the solution of equation \eqref{eq1} with 
potential \eqref{eq18} from the solution 
${\it Y_s}={\frac {1}{\sqrt {{r}^{2}+{z}^{2}}}}$ with potential $u=0$.
Let us consider 
${\it Y_h}={r}^{2}-2\,{z}^{2}, { \it Y}={\it Y_s}$.
From the equations \eqref{eq15}, \eqref{eq16} we
obtain the solution of equation \eqref{eq1} with potential
 \eqref{eq17}
\begin{equation} \label{eq19}
{\tilde {Y}}_{s} =
{\frac {rz}{ \left( {r}^{2}-2\,{z}^{2} \right) \sqrt {{r}^{2}+{z}^{2}}
}}
\end{equation}
Then  consider 
$ {\it Y_h}={\tilde {Y}}_{1}, { \it Y}={\tilde {Y}}_{s}$
and get from the equations \eqref{eq15}, \eqref{eq16} 
the following solution of equation \eqref{eq1} with potential
 \eqref{eq18}
\begin{equation} \label{eq20}
{\tilde {\tilde {Y}}}_{s} =
{\frac {3\,{r}^{4}-C_1}{\sqrt {{r}^{2}+{z}^{2}} \left( 4\,{r}^{2}{z}^{2}
+{r}^{4}+C_1 \right) }}
\end{equation}
The integration constant arising is set to zero.

In the case ${\it u} < 0$ equation \eqref{eq1} coincides with 
Helmholtz equation for axially symmetric medium.  For the second 
example of  of the generalized Moutard transformation 
application let us consider 
$u=-{k}^{2}, {\it Y_h}=\sin \left( kz \right), 
{ \it Y}=\cos \left( kz \right)$.
From the equations \eqref{eq11}, \eqref{eq15}, \eqref{eq16} we
obtain potencial
\begin{equation} \label{eq21}
{\tilde {u}}_{2}   =
-{k}^{2}+{\frac {1}{{r}^{2}}}+{\frac {2\,{k}^{2}}{ \left( \sin \left( kz \right) 
 \right) ^{2}}}
\end{equation}
and solution of equation \eqref{eq1} with this potential
\begin{equation*}
{\tilde {Y}}_{2} =
{\frac {{r}^{2}+{\it C_2}}{r\sin \left( kz \right) }}
\end{equation*}
where $C_2$ is an arbitrary constant. 
Let us make the second 
generalized Moutard transformation. Consider
$u={\tilde {u}}_{2}, {\it Y_h}={\tilde {Y}}_{2}$.
From the equation \eqref{eq11} we obtain potencial 
\begin{equation} \label{eq22}
{\tilde {\tilde {u}}}_{2}  =
-{k}^{2}+{\frac {4}{\left( {r}^{2}+{\it C_2} \right)}}-{\frac {8\,{\it 
C_2}}{ \left( {r}^{2}+{\it C_2} \right) ^{2}}}
\, .
\end{equation}
This potential evidently satisfies  
${\tilde {\tilde {u}}}_{2}< 0$  if $C_2 > 4\,{k}^{-2}$.
The simple example of solution ${\left(r\,{\it Y_h} \right)}^{-1}$
for potencial \eqref{eq22} is 
\begin{equation*}
{\tilde {\tilde {Y}}}_{2} =
{\frac {\sin \left( kz \right) }{{r}^{2}+{\it C_2}}} \,.
\end{equation*}

The differential operator in  \eqref{eq1} is invariant
under shift of $z$. So $z$ can be replaced by $z+z_0$
in all formulae for potencials and solutions presented above.
For example consider $u=-{k}^{2}$ and  
\begin{equation*}
 {\it Y_h}={\frac {\sin \left( k\sqrt {{r}^{2}+ \left( z+{\it z_0} \right) ^{2}}
 \right) }{\sqrt {{r}^{2}+ \left( z+{\it z_0} \right) ^{2}}}},
{ \it Y}={\frac {\cos \left( k\sqrt {{r}^{2}+ \left( z+{\it z_0} \right) ^{2}}
 \right) }{\sqrt {{r}^{2}+ \left( z+{\it z_0} \right) ^{2}}}}
\,.
\end{equation*}
From the equations \eqref{eq11}, \eqref{eq15}, \eqref{eq16} we
obtain potencial
\begin{multline} \label{eq23}
{\tilde {u}}_{3}   =
-{k}^{2}+{\frac {1}{{r}^{2}}}+{\frac {2\,{k}^{2}}{ \left( \sin \left( k\sqrt {{r}^{2}+
 \left( z+{\it z_0} \right) ^{2}} \right)  \right) ^{2}}}
\\
-{\frac {2\,k
\cot \left( k\sqrt {{r}^{2}+ \left( z+{\it z_0} \right) ^{2}}
 \right) }{\sqrt {{r}^{2}+ \left( z+{\it z_0} \right) ^{2}}}}
\end{multline}
and solution of equation \eqref{eq1} with this potential
\begin{equation*}
{\tilde {Y}}_{3} =
{\frac {z+{\it z_0}+C_3\,\sqrt {{r}^{2}+ \left( z+{\it z_0} \right) ^{2}}
}{\sin \left( k\sqrt {{r}^{2}+ \left( z+{\it z_0} \right) ^{2}}
 \right) r}}
\end{equation*}
where $C_3$ is an arbitrary constant. 
Let us make the second 
generalized Moutard transformation. Consider
$u={\tilde {u}}_{3}, {\it Y_h}={\tilde {Y}}_{3}$.
From the equation \eqref{eq11} we obtain potencial 
\begin{multline} \label{eq24}
{\tilde {\tilde {u}}}_{3}  =
-{k}^{2}+2\, \left( z+{\it z_0}+C_3 \,\sqrt {{r}^{2}+ \left( z+{\it z_0}
 \right) ^{2}}\right) ^{-2}
\\
+{\frac {2\,C_3 \left( z+{\it z_0}
 \right) }{\sqrt {{r}^{2}
+ \left( z+{\it z_0} \right) ^{2}} \left( z+
{\it z_0}+C_3\,\sqrt {{r}^{2}+ \left( z+{\it z_0} \right) ^{2}} \right) 
^{2}}}
\, .
\end{multline}
This potential for $z\ge 0$ satisfies  
${\tilde {\tilde {u}}}_{3}< 0$  if ${\frac {2}{{{\it z_0}}^{2} \left( 1+C_3 \right) }}<{k}^{2}$.
The simple example of solution ${\left(r\,{\it Y_h} \right)}^{-1}$
for potencial \eqref{eq24} is 
\begin{equation*}
{\tilde {\tilde {Y}}}_{3} =
{\frac {\sin \left( k\sqrt {{r}^{2}+ \left( z+{\it z_0} \right) ^{2}}
 \right) }{z+{\it z_0}+C_3\,\sqrt {{r}^{2}+ \left( z+{\it z_0} \right) ^{
2}}}}
 \,.
\end{equation*}

\section{Results and Discussion}

The stationary Schr\"odinger equation 
in the case of axial symmetry is investigated.
Using the approach of the papers 
\cite {Kudryavtsev2013}, \cite {Kudryavtsev2016}
the nonlocal Darboux transformation of the two - dimensional
stationary Schr\"odinger equation in cylindrical coordinates  
is considered. It is shown that a special case of 
the nonlocal Darboux transformation provides the generalization of 
the Moutard transformation. Formulae for the generalized 
Moutard transformation are obtained.  New examples of two - dimensional  
potencials and exact solutions for the stationary axially symmetric 
Schr\"odinger equation are obtained as an application of 
the generalized Moutard transformation.

The generalized Moutard transformation
can be initiated by any exact solution and applied repeatedly.
Thus further application of the generalized Moutard transformation
can provide a lot of new examples of potencials and exact solutions 
for the stationary axially symmetric Schr\"odinger equation.

\end{document}